\documentclass[dvips,12pt]{article}
\usepackage{graphicx}
\begin{document} 
\thispagestyle{empty} 

\begin{center}
{\large{\bf Critical Opalescence in Baryonic QCD Matter}}\\
\vspace{2cm} 
{\large N.~G.~Antoniou, F.~K.~Diakonos, A.~S.~Kapoyannis}\\ 
{\it Department of Physics, University of Athens, 15771 Athens, Greece}\\
\vspace{0.5cm}
{\large K.~S.~Kousouris}\\
{\it National Research Center ``Demokritos'', Institute of Nuclear 
Physics, Ag. Paraskevi, 15310 Athens, Greece}
\vspace{1cm}

\end{center}
\vspace{0.5cm}

\begin{abstract}
We show that critical opalescence, a clear signature of second-order phase transition in conventional 
matter, manifests itself as critical intermittency in QCD matter produced in 
experiments with nuclei. This behaviour is revealed in transverse momentum spectra 
as a pattern of power laws in factorial moments, to all orders, associated with baryon 
production. This phenomenon together with a similar effect in the isoscalar sector of 
pions (sigma mode) provide us with a set of observables associated with the 
search for the QCD critical point in experiments with nuclei at high energies.
\end{abstract}

\vspace{3cm}

\vspace{1cm} 

{\small
{\it e-mail addresses:} 

nantonio@phys.uoa.gr (N.G.~Antoniou)

fdiakono@phys.uoa.gr (F.K.~Diakonos)

akapog@phys.uoa.gr (A.S.~Kapoyannis)

kousouris@inp.demokritos.gr (K.S.~Kousouris)}

\newpage
\setcounter{page}{1}

It has been suggested recently that the baryon number fluctuations, in
event-by-event studies of multiparticle production, may lead to new
observables, in experiments with nuclei, associated with the existence and
location of the critical point in the QCD phase diagram [1]. Theoretically,
the significance of the baryon number density $n_B(\vec{x})$ near the
critical point comes from the fact that its fluctuations, expressed in terms
of the density-density correlator $\left<n_B(\vec{x})n_B(0)\right>$, follow
the same power law as the sigma field correlator
$\left<\sigma(\vec{x})\sigma(0)\right>$
associated with the chiral order parameter
$\sigma(\vec{x})\sim\bar{\Psi}\Psi$
[1,2]. As a result, the critical thermodynamics of QCD matter in equilibrium 
can be
formulated in terms of the baryon number density $n_B(\vec{x})$,
an alternative but equivalent order parameter related to the final state of
nuclear collisions when they freeze out close to the critical point. The
corresponding theory of critical power laws in the baryon sector, the observability of
which we explore in this Letter, is based on the effective action belonging to $3d$
Ising universality class:
\begin{equation}
\Gamma_c\left[n_B\right]=T_c^{-5}g^2\int d^3\vec{x}
\left[\frac{1}{2}\left|\nabla n_B\right|^2+
Gg^{\delta-1}T_c^8 \left|T_c^{-3}n_B\right|^{\delta+1}\right]
\end{equation}
where $\delta\simeq5$ is the isotherm critical exponent and $G$ a universal,
dimensionless coupling in the effective potential, with a value in the range
$G\simeq1.5-2$ [3]. In writing eq.~(1) we have considered, following the above
discussion, the order parameter $m(\vec{x})= gT_c^{-2}n_B\left(\vec{x}\right)$,
associated with the fluctuations of baryon number density at the critical point;
the factor $T_c^{-2}$ guarantees dimensional consistency and $g$ is a nonuniversal
dimensionless constant.

The free energy (1) in order to describe a critical system of QCD matter
produced in nuclear collisions, must be adapted to the relativistic geometry
of the collision. To this end, the longitudinal coordinate $x_{\parallel}$ is
replaced by the space-time rapidity $y$ and the corresponding measure of
integration in (1) takes the form: $dx_{\parallel}=\tau_c\cosh y dy$ where
$\tau_c$ is the formation time of the critical point. In the central region
of size $\delta y$ the configurations $n_B\left(y,\vec{x}_{\perp}\right)$
contributing to the partition function are boost invariant quantities
defining at the same time two dimensional baryon-number density
configurations $\rho_B\left(\vec{x}_{\perp}\right)$ in the transverse plane:
$n_B\left(\vec{x}_{\perp}\right)=\rho_B\left(\vec{x}_{\perp}\right)
\left[2\tau_c\sinh\left(\delta y/2\right)\right]^{-1}$.
Integrating now in rapidity and rescaling the basic variables:
$\vec{x}_{\perp} \rightarrow T_c\vec{x}_{\perp}$,
$\rho_B \rightarrow T_c^{-2}\rho_B$, eq.~(1) is simplified as follows:
\begin{equation}
\Gamma_c\left[\rho_B\right]=Cg^2\int d^2\vec{x}_{\perp}
\left[\frac{1}{2}\left|\nabla_{\perp} \rho_B\right|^2+
G\left(gC\right)^{\delta-1}\rho_B^{\delta+1}\right]
\end{equation}
\[
C\equiv\left(T_c \tau_{eff} \right)^{-1} \;,\;
\tau_{eff} \equiv 2\tau_c \sinh \left(\delta y/2\right)
\]
The critical fluctuations of systems belonging to the class (2) of a $2d$
effective action can be consistently described in a scheme where the
saturation of the partition function is obtained considering the contribution
of the singular solutions $\rho_B^{\left( s \right)}$ of the Euler-Lagrange
equation [4]:
\begin{equation}
Z\simeq \sum_s e^{-\Gamma_c [\rho_B^{\left( s \right)}]} \; ; \;
\nabla_{\perp}^2 \rho_B^{\left( s \right)} -
\left(\delta+1\right) G \left(gC\right)^{\delta-1}
\left[\rho_B^{\left( s \right)}\right]^{\delta}=0
\end{equation}
The description of the critical system in this scheme is optimal if
$Cg^2 \gg 1$.

The main characteristics of critical QCD matter ($\delta \simeq 5$,
$G\simeq 2$) produced in nuclear collisions and described by eqs.~(3) are
summarized as follows [4]:
\newline (a) The system is organized within critical domains (clusters) of a
maximal size (correlation length):
$\xi_{\perp}\simeq \frac{\pi\tau_c}{8}\sinh\left(\delta y/2\right)$.
\newline (b) For typical nuclear collisions ($\tau_c\simeq 10$ fm,
$\delta y \geq 2$) the transverse correlation length becomes sufficiently
large ($\xi_{\perp}\geq 6$ fm) allowing for critical fluctuations to
develop in full strength when the system freezes out near the critical point.
\newline (c) Within a critical domain
($\left|\delta \vec{x}_{\perp}\right| \leq \xi_{\perp}$) the correlator
obeys a power law corresponding to a fractal dimension
$d_F=\frac{2\delta}{\delta +1}$ ($d_F\simeq\frac{5}{3}$):
$\left<\rho_B(\vec{x}_{\perp})\rho_B(0)\right>\sim
\left|\vec{x}_{\perp}\right|^{d_F -2}$.
This power law is the origin of critical opalescence [5] in baryonic QCD
matter produced in high-energy nuclear collisions.

In conventional matter (QED matter) the phenomenon of critical opalescence
is revealed when light of long wavelength (comparable to the correlation
length) scatters from a substance near criticality. The intensity of
scattered light is proportional to the Fourier transform of the
density-density correlator and becomes singular for small momentum transfer
$k$ giving rise to a macroscopic effect [5]. In QCD matter, correspondingly,
the Fourier transform of the correlator
$\left<\rho_B(\vec{x}_{\perp})\rho_B(0)\right>$ obeys also a power law:
\[
\left<\rho_{\vec{k}}\rho_{\vec{k}}^*\right> \sim
\int d^2 \vec{x}_{\perp} e^{-i \vec{k}\cdot\vec{x}_{\perp}}
\left<\rho_B(\vec{x}_{\perp})\rho_B(0)\right>
\]
\begin{equation}
\left<\rho_B(\vec{x}_{\perp})\rho_B(0)\right> \sim
\left|\vec{x}_{\perp}\right|^{d_F-2} \; ; \;
\left|\vec{x}_{\perp}\right|\stackrel{<}{\sim}\xi_{\perp}
\end{equation}
\[
\left<\rho_{\vec{k}}\rho_{\vec{k}}^*\right> \sim
|\vec{k}|^{-d_F} \; ; \;
|\vec{k}| \stackrel{>}{\sim} \xi_{\perp}^{-1}
\]
where $\rho_{\vec{k}}$ is the Fourier transform of $\rho_B(\vec{x}_{\perp})$
and $\left<\rho_{\vec{k}}\rho_{\vec{k}}^*\right>$ the two-particle
correlator in transverse momentum plane, associated with baryon production in
nuclear collisions. More precisely,
\begin{equation}
\left<\rho_{\vec{k}}\rho_{\vec{k}}^*\right> =
\sum_{\vec{p}} \left<\rho_B(\vec{p})\rho_B(\vec{p}+\vec{k})\right> \sim
\left<\rho_B(0)\rho_B(\vec{k})\right>
\end{equation}
under the assumption that the dependence of 
$\left<\rho_B(\vec{p})\rho_B(\vec{p}+\vec{k})\right>$
on the reference momentum $\vec{p}$, is weak.

The power law (4) in momentum space can be generalized to a self-similarity
property of multiparticle correlators, valid for the solution (3) of QCD
matter at criticality [4]:
\begin{equation}
\left<\rho_B(\lambda\vec{k}_1)\cdots\rho_B(\lambda\vec{k}_{q-1})\rho_B(0)\right>
=\lambda^{-d_F(q-1)}
\left<\rho_B(\vec{k}_1)\cdots\rho_B(\vec{k}_{q-1})\rho_B(0)\right>,\;
(q=2,3,\ldots)
\end{equation}
In geometrical language, eqs.~(6) define a fractal dimension $\tilde{d}_F$ in
transverse momentum space, independent of $q$ ($\tilde{d}_F=2-d_F$),
directly related to the isotherm critical exponent of QCD:
$\tilde{d}_F=\frac{2}{\delta+1}$ [6]. The observable effect implied by (6) is
a specific class of power laws, satisfied by the scaled factorial moments of
all orders [7]:
\begin{equation}
F_q(M)=
\left[\frac{\left<N_B(N_B-1)\cdots(N_B-q+1)\right>}{\left<N_B\right>^q}\right]_
{\delta S_M} \sim
M^{d_F (q-1)}\;\; (M \gg 1)
\end{equation}
($q=2,3\ldots$) defined in small domains $\delta S_M$ of transverse momentum
plane, constructed by a subdivision of the available space in $M^2$
two-dimensional cells ($\delta S_M \sim M^{-2}$). The power laws (7) with
a characteristic linear spectrum of indices
$\alpha_q=(2-\tilde{d}_F)(q-1)$, describe, in general, the fluctuations in a
second-order phase transition [6,8] but here, in particular, they reveal the
effect of critical intermittency associated with the production of baryonic
QCD matter near the critical point. This phenomenon is the analogue of
critical opalescence, observed in conventional matter near criticality. Both
phenomena have their origin in the power laws of the correlators in
configuration space, implying giant density fluctuations at the critical
point as a result of the appropriate fractal geometry, valid within the
universality class of the critical system under consideration.

In order to make the prediction (7) a transparent signature for the QCD
critical point in experiments with nuclei, one may proceed to a Monte Carlo
simulation of critical events in the case of a typical process A+A at high
energies. We have chosen collisions of medium size nuclei producing net baryons in
the central region with average multiplicity $\left<N_B\right>\simeq 60$, at
energies corresponding to CERN-SPS. At the critical point we expect an
extended plateau in the rapidity spectrum of net baryons [1] allowing for a
choice $\delta y \simeq 5$ in the system under consideration. A
self-consistency requirement in this treatment is the constraint $Cg^2~\gg~1$
which may justify a posteriori the saddle point saturation of the partition
function leading to eqs.~(3). In order to clarify this issue for the above
system under simulation, we consider the following generic form coming from
the thermal average $\left<\rho_B\right>$ with the aid of eqs.~(3):
\begin{equation}
\left<N_B\right>\simeq  \frac{1}{g} \left(\frac{S_{cr}}{CG^{1/\delta}}\right)^
{\frac{\delta}{\delta+1}}
\frac{\Gamma\left(\frac{2}{1+\delta}\right)}
{\Gamma\left(\frac{1}{1+\delta}\right)}\;\;\;
(\delta\simeq5,\;G\simeq2,\;\tau_c\simeq 10 fm)
\end{equation}
where $S_{cr}$ is the area of the critical cluster,
$S_{cr}\simeq {\cal O}(\xi^2_{\perp})$.
For the specifications of the system under study we find
$Cg^2~\simeq ~{\cal O}(10^3)$,
a result which fulfils comfortably the consistency requirement ($Cg^2~\gg~1$).

The simulation of events involving critical fluctuations in the baryonic density
requires the generation of baryon transverse momenta correlated according to the
power-law (4).
One possibility to produce such a set of momenta is to use the method of L\'{e}vy
random walks [9]. It is convenient to perform independent one-dimensional walks in 
each transverse momentum component separately and then form the Cartesian product in 
order to obtain the corresponding vectors. In this case the successive steps in each
dimension are chosen according to the probability density:
\begin{equation}
\tilde{\rho}(p_i)=\frac{\nu p_{min}^{\nu}}{1-\left(\frac{p_{min}}
{p_{max}}\right)^{\nu}} p_i^{-1 -\nu}~~~~;~~~~p_{min} \leq p_i \leq p_{max},
~~~~i=x,y 
\end{equation}
To generate a single event one performs $n-1$ random walk steps in each direction ($p_x$
or $p_y$) where $n$ is the multiplicity of the event. At each step the updated position 
of the walker determines the $p_x$ (or $p_y$) coordinate of the baryon transverse
momentum respectively. The starting transverse momentum vector in each event is chosen 
uniformly in $[-p_{max},p_{max}] \times [-p_{max},p_{max}]$. The produced set of baryon transverse 
momenta possesses the correct fractal dimension $\tilde{d}_F=\frac{1}{3}$ provided that 
$\frac{p_{min}}{p_{max}} \approx 10^7$ and $\nu=\frac{1}{6}$. 
The described algorithm can be used to generate an ensemble of events involving critical baryon density
fluctuations. The corresponding fractal pattern can be revealed through factorial moment analysis [7].
In Fig.~1 we show, in a log-log plot, the results obtained for an ensemble of 100
critical events each one having the multiplicity $n=\langle N_B \rangle =60$. 
The full circles are the calculated moments up to the 4th order. The solid lines are the theoretical predictions according to eq.~(7). It is clearly
seen that the factorial moments follow to a good approximation and for a wide range of scales a power-law behaviour $F_q \sim (M^2)^{s_q}$ with exponents 
$s_q$ satisfying very well the theoretical prediction $s_q=(q-1)(1-\frac{\tilde{d}_F}{2})$. In the same plot we show for comparison the behaviour of the second 
moment (grey triangles) in a conventional system corresponding to the formation of mixed events from the original 
ensemble of the critical events. As expected, the effect of critical fluctuations has
disappeared in the system of mixed events (the slope $s_2^{(m)}\simeq0$). 

\vspace{1cm}
{\includegraphics[width=6in,height=4in]{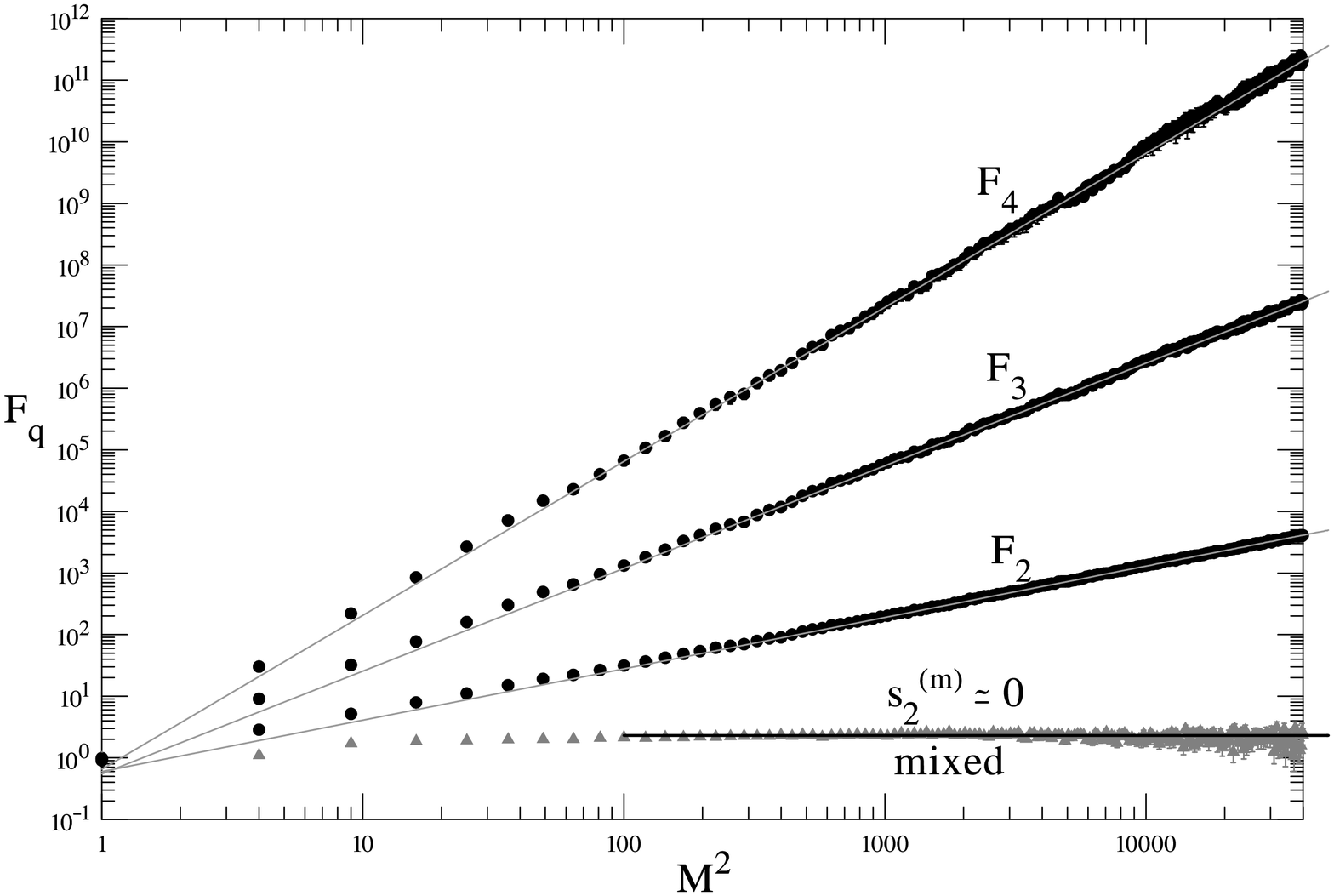}}

{\bf Fig.~1.} {\it The log-log plot of the factorial moments of order $q=2,3,4$ (full circles) for an ensemble of 100 critical events 
generated by the Monte-Carlo algorithm described in the text. The corresponding theoretical power-law predictions are shown 
with solid lines. The grey triangles present the result for the second moment for a set of mixed events.}
\vspace{1cm}

We have shown that baryon-number density fluctuations, near the QCD critical
point, develop a distinct pattern of power laws, in transverse momentum
plane, associated with the effect of critical intermittency in QCD matter.
The observability of this effect is enhanced by the conjecture that the same
power laws (Fig.~1) are expected also in the case of net proton-number density
fluctuations, avoiding therefore the observational ambiguities from the
contribution of neutrons to the factorial moments (7). In fact, it has been
argued by Hatta and Stephanov [1] that the singular parts of baryon
susceptibility ($\chi_B$) and proton number fluctuations coincide,
$\chi_B \sim \left<\delta N_{p-\bar{p}} \delta N_{p-\bar{p}}\right>$.
As a result, the power laws associated with these singularities are the same
for both the baryon-number and proton-number density correlators. With these
remarks we may
conclude that proton number measurements in transverse momentum plane may
reveal the effect of critical baryon-number intermittency in nuclear
collisions, as a signal of QCD criticality. If one combines this observable
effect with the corresponding phenomenon in the sigma mode of pion production
[10] then a complete set of observables, associated with the existence and
location of the QCD critical point, naturally emerges.

\vspace{1cm}
This work was supported in part by the Research Committee of the University of Athens
(research funding program ``Kapodistrias'') and the EPEAEK research funding program ``PYTHAGORAS I'' (70/3/7315) (Ministry of Education).

\end{document}